# On the FP7 BlackSeaHazNet Project and Its Possible Application for Harmonic Existence of the Regions

**Strachimir Chterev Mavrodiev [a],*, Lazo Pekevski [b]**

[a] Institute for Nuclear Research and Nuclear Energy, Bulgarian Academy of Sciences, Bul. Tzarigradsko shose 72, Sofia 1784, Bulgaria.
[b] Seismological Observatory, Faculty of Natural Sciences and Mathematics, Sts. Cyril and Methodius University, Skopje, Macedonia.

## ABSTRACT

This paper describes breifly our long journey from the time of working for "Clean and peaceful Black Sea" in the 80-th to the FP7 support of the project "Complex research of Earthquake's Forecasting Possibilities, Seismicity and Climate Change Correlations". The participants are from Slovenia, Macedonia, Greece, Turkey, Bulgaria, Georgia and Armenia. The main aim of the project and the preliminary results were published in the Ohrid, Macedonia (ISBN 978-9989-631-04-7) and Tbilisi, Georgia (UDC 550.348.436+551.583.C-73, ISSN 2233-3681), also the proceedings are presented.

## 1. Introduction.

The full name of the project is FP7, Marie Curie Actions, People, International Research Staff Exchange Scheme, Scientific panel: Environment, Acronym: BlackSeaHazNet, No 246874, duration of 24 months starting in 2011.

*Corresponding author, Tel.: 02 9795621, Fax: +359 2 9753619, Mobile: +359 898538006
E-mail addresses: schtmavr@yahoo.com, lazopekevski@hahoo.com

**Table 1.** List of Partner Organisations

| Partner Number | Partner name | Partner short name | Country |
|---|---|---|---|
| **1** *Beneficiary* 1 | Institute for Nuclear Research and Nuclear Energy, Bulgarian Academy of Sciences | INRNE BAS | Bulgaria |
| **2** *Beneficiary* 2 | National Institute for Geophysics, Geodesy and Geography, Bulgarian Academy of Sciences | NIGGG BAS | Bulgaria |
| **3** *Beneficiary* 3 | Institute of Oceanography , Hellenic Center for Marine Research | HCMR | Greece |
| **4** *Beneficiary* 4 | Josef Stefan Institute | JSI | Slovenia |
| **5** *Beneficiary* 5 | Karts Research Institute, Znanstvenoraziskovalni Center Slovenske Akademije Znanosti in Umetnosti | SRC SASA SI | Slovenia |
| **6** *Beneficiary* 6 | Geologicheski Institut pri BAN St. Dimitrov | GI BAS | Bulgaria |
| **7** *Beneficiary* 7 | Space and Solar-Terrestrial Research Institute | SSTRI BAS | Bulgaria |
| **8** *Beneficiary* 8 | Ss. Cyril and Methodius University in Skopje | SORM UKIM | FYRO Macedonia |
| **9** *Beneficiary* 9 | Earth and Marine Sciences Institute, TUBITAK Marmara Research Center | TUBITAK MAM | Turkey |
| **10** Partner 1 | Western Survey for Seismic Protection SNCO, Ministry of Emergency Situations | WSSP | Armenia |
| **11** Partner3 | Mikheil Nodia Institute of Geophysics, Ministry of Education and Science of Georgia | GI MES | Georgia |
| **12** Partner 4 | Ilia State University*ILIAUNI | ISU | Georgia |
| **13** Partner 5 | Geophysics Institute named after S.I.Subbotin of the National Academy of Sciences in Ukraine | IG NASU | Ukraine |
| **14** Partner 6 | National Antarctic Scientific Center | NASC | Ukraine |
| **15** Partner 7 | Odessa National Polytechnic University | ONPU | Ukraine |
| **16** Partner 8 | Institute for Nuclear Research of NAS of Ukraine | INR NASU | Ukraine |

## 2. Project summary

## 2.1. Balkan, Black Sea, Caucasus, Caspian NETWORK for Complex Research of Earthquake's forecasting Possibilities, Seismicity and Climate Change Correlations

The purpose of the project, BlackSeaHazNet, is a development of long-term research cooperation through coordinated joint program for exchange of data, know-how and scientists. The partnership in experimental and theoretical aspects of geophysics will be focused on a creation of fundamentals of a Complex Program for investigation of the possibilities to forecast earthquake's time, hypocenter magnitude, and intensity using reliable precursors. For this aim, a monitoring of the following parameters is envisaged:

- geophysical and seismological variability over South-East Europe;

- water sources and their Radon, Helium, and other gases concentrations;

- crust temperature;

- electromagnetic field variations under, on and above Earth's surface;

- satellite monitoring of meteorological parameters, including earthquake clouds;

- electrical charge distributions and variations in ionospheric parameters.

- near-space monitoring, aimed to detect and exclude the external influences (i.e., those of Sun and Interplanetary medium variations, Cosmic rays) from the meaningful signal of the solid Earth.

- biological precursors.

For many variables, well working global and regional monitoring exists (for example INTERMAGNET monitoring). For others (for example: i.e., monitoring of Earth's currents distribution) monitoring have to be created.

The complex monitoring of the broad variety of parameters defines the output of the Program including: i) estimations of different time scales for more clearer understanding of the Earth's system natural variability, ii) risk assessment of the appearance of hazards for society events related to earthquakes, climatic changes, etc., and iii) people's response to an abrupt change in the monitored parameters.

The proposed regional network can be considered as a first step in creation of a wide interdisciplinary scientific consortium capable of formulation of a more adequate paradigm of climate variability and climatic change (distinguishing the differences between them), Earth seismic processes and the actual problem of their forecast. If the hypothesis for Georeactors (Rusov et al, 2006, 2009, 2010; Feoktistov, 1998; 238U; Teller, 1996; 232Th type) net, as possible reason for Climate change will be confirmed, the surprising new knowledge for Climate variations and its connection with Earth's seismicity can be obtained. Conformation of the hypothesis for existence of new type self-regulated nuclear reactors will enhance the physical understanding of different time scales of Climate - Climate and Climate - Weather transitions and Climate - Seismicity correlations.

The exchange of people, methods for data analysis, different approaches, models, etc., is expected to improve the current knowledge and to build up new concepts, models, etc. Results will be reported and discussed in the workshops and regional training seminars following the principle of brain-storming.

Establishment of such a multidisciplinary program is supported by every partner institutions affording an opportunity to use their own research resources.

## 2.2. Quality of the Exchange Programme

### 2.2.1. Objectives and relevance of the joint exchange program

The main measurable goals of BlackSeaHazNet programme are:

- To establish real-time closer collaboration between partners for every experimental, theoretical tasks.
- For easier real time exchange of results and for education of the wide public, including authorities, to develop the website Earthquake Forecast Using Reliable Earthquake Precursors (http://theo.inrne.bas.bg/~mavrodi/) to the website named Complex Research of Earthquake's Forecasting Possibilities, Seismicity and Climate Change Correlations, where the project BlackSeaHazNet will be presented task by task, the results of regional monitoring, scientific results, materials from workshops and training seminars;
- To formulate the theoretical need and technical description for new variables monitoring for the production experimental data, for successful work in step by step solutions, the problems from Complex Research of Regional Earthquake's Forecasting Possibilities, Seismicity and Climate Change Correlations and reasons for their variations;
- To summarise exiting regional real time and history data, to analyse their quality through inventory of their acquisition systems for preliminary archiving;
- To compare the existing methods for visualization and analysis of the data, estimation of the risk and forecast of earthquakes (if any), which may be utilized by all participants in the near future;l
- To create a regional network for continuous monitoring of the earthquakes' precursors;
- To integrate and synthesise the existing research results related to variations of seismic activity and climatic system on different time scales - for better estimation and predictions of forecasting their future changes;

- To promote the joint research activities on regional and European levels, involving Europe's near neighbour countries into the ad hoc research activities;
- To transfer the "pure scientific" results into a knowledge for the society, facilitating the two-way communication processes with policy and general public for better understanding of the natural hazard risks, and more effective mitigation of its influence on people's life and environment;
- To establish financial support for the realizing of the wide interdisciplinary experimental and theoretical work for solving the imminent when, where and how earthquake's forecast as well as the derived results for different time scale estimations for some natural hazards and risk assessment, and the correlations between human behaviour and electromagnetic, and other geophysical variables variations in the different variants of the real working Complex Research of Regional Earthquake's Forecasting, Seismicity and Climate Change variations.

## 2.2.2. Complementary / synergies between partners

The degree of integration, which BlackSeaHazNet program aims to expand and deeper based on long lasting traditional bi- and multi-lateral collaborations among many of the partners.

A great part of participants are involved in the Balkan-South-Caucasian network for continuous monitoring of solar and geophysical parameters. Moreover, bi-lateral agreements exist between most of countries attending the BlackSeaHazNet program. The national support, provided over the years to the partners, is a proof of interest of their national funding organisations for international cooperative efforts to understand and predict the risk of natural hazardous events. Thus funding of this project, by IRSES scheme of FP7 framework program, will strengthen the existing networks trough provision of new activities, which foster the further integration of the South-East European community. Two types of activities are planned during the duration of the Program:

### 2.2.3. Integrating activities

### 2.2.3.1. International multi-disciplinary coordination and collaboration

The main benefits from this multi-disciplinary networking include: improved efficiency and cost-effectiveness of scientific research, expand the boundaries of existing research for building up a *new cutting-edge knowledge;* the possibility to find a solution of unresolved till now problems due to the limitations of the existing theories, models, etc., as well as enhance the possibility to influence the public policies affecting thus far the quality of life.

### 2.2.3.2. Access to information and research infrastructure

All project participants will have the possibility to exchange their own data or to use standardized databases to synchronize the used approaches, to share their research infrastructure, etc. The synergy of research programs between EU countries and near neighbour countries allowing a diffuse of information and different databases may be extremely beneficial for solving such complicated problems.

### 2.2.3.3. BlackSeaHazNet web-page

*BlackSeaHazNet web-page* will maintain communications between participants, providing a tribune for exchange of ideas, data, results, as well as including the step by step real time common monitoring data.

### 2.2.3.4. Joint research activities

Reevaluation of the existing models of Earth's crust conditions, Earth and atmosphere variables, solar and space influences, in order to better understand the physics of Earthquakes; systematization of the earthquake parameters (magnitude, seismic moment, intensity, depth, size of volume and surface fault), and earthquake occurrence by applying different statistical methods, including non-linear and inverse problem methods;

Statistical analysis of empirically determined relations between different geophysical parameters through non-linear inverse problem methods, including special studies of aftershocks activity;

Analysis of the factors influencing climate variability aimed to answer the question: Is the currently observed global warming a manifestation of natural climate fluctuations or it is an indication of an increasing instability of the contemporary climate due to the increased anthropogenic forcing; and investigation of possible relations between climatic variations and seismic activity;

The strong integration within the BlackSeaHazNet program is also demonstrated trough interlinking and interdependence of all tasks and activities of the network. The horizontal link between partners aims to integrate the existing knowledge about the processes governing variations of solid Earth parameters, in order to build a physical model of earthquakes. Most partners are involved in multiple working packages within the project, which favour their effective integration within BlackSeaHazNet program. Most tasks of BlackSeaHazNet project have by their own nature, meaning of integration between the participants illustrated by the common use of data bases, sharing of facilities, exchange of researchers, etc. The organisation of a series of five workshops is aimed to mobilize the interdisciplinary skills and the expertise of the project participants for gathering the optimal conditions for solution of the problems, including complex programming.

BlackSeaHazNet network will make an extensive use of Internet-based techniques to foster communication within scientific community and to implement all BlackSeaHazNet activities. In this context the BlackSeaHazNet web-page (based on http://stardust.inrne.bas.bg/mavrodi/) devoted on complex research of earthquake's forecasting possibilities, seismicity and climate change relations can be considered as the overall integrating toll of the Network.

## 2.2.4. Transfer of Knowledge

### 2.2.4.1. Quality and mutual benefit of the transfer of knowledge

Knowledge Transfer is the two-way flow of ideas, expertise and people between the research base and wide users in society. A special stress is put on the exchange of knowledge in the the BlackSeaHazNet program - demonstrated by the definition of a special work-package devoted on this issue.

One of the main benefits of BlackSeaHazNet realization will increase cohesion between scientists in different branches of geophysics, nuclear physics, etc. It will provide an opportunity for them to develop new skills and collaborations, to capitalise on their research knowledge and findings beyond their narrow areas of high quality expertise.

An interactive transfer of data, ideas, specific approaches in data analysis and modelling between project participants and with international scientific communities - through a direct exchange of people, series of workshops, lectures and trainings for younger staff - is a good prerequisite for building of new knowledge. Jointly executed research has a potential to inspire the experienced scientist to develop strategies for knowledge transfer, as well as and to tailor them to potential users and society.

## 2.2.4.2. Adequacy and role of staff exchange with respect to the transfer of knowledge

Effective knowledge transfer strategies rely on the capacity of Universities and Academic institutions to shape their knowledge transfer approaches and activities, in partnership with their various communities, and to respond creatively to the distinctive needs of those communities. From this perspective, a "healthy" system of knowledge transfer should demonstrate considerable diversity in knowledge transfer approaches and activities, both *within and across institutions, and across disciplines and national research priorities*. In this context, the possibility for exchange of scientists within IRSES can be a great step forward in the difficulty of transfering knowledge between project participants. The exchange of ideas during the series of seminars will be extended logically to a two way flow of data, methods and approaches for data monitoring and analysis, for creation of new or improvement of the existed approaches stimulated by the multi-disciplinary brainstorming, for reassessment of the existed models with real possibilities for their improvement via comparison with experiment, and for ability to take advantage of scientific infrastructure of the host country.

As a final accord, it should be reminded that the "innovation does not necessarily require new technologies, existing technologies may be applied in new ways" resulting in a tremendous improvement of the existing or creation of qualitatively new knowledge (FP7, basic ideas).

The effective transfer of knowledge is guaranteed from the visits of exchange programs, attended from seminars and lectures on the topic of the visit, the Programs of workshops and by the fact, that materials will be published in the end of every Workshop, as well as in the Website and will be distributed between authorities.

## 3. History

### 3.1. Hypothesis

The hypothesis for possible correlations between the earthquakes, the magnetic fields, Earth's horizontal and vertical currents in the atmosphere, was born when in early 1988, the historical data on the Black Sea was systemized. The fire pillars, observed during the Crimean earthquake in 1928, could be explained with the dehydration of the biogas condensated at the time of the spread out of the earthquake's waves in the sediments, and the ignition, which follows in the presence of a high degree of ionization of the atmosphere. Such currents can be measured, for example through a precise measurement of the Earth's magnetic fields. A common model for the origin of the attendant Earth's currents before, during and after an earthquake does not exist.

### 3.2. One projection of geomagnetic field monitoring

In December 1989, a continuous measurement of a projection of the Earth's magnetic field (F) with a magnetometer (know-how of JINR, Dubna, Boris Vasiliev) with absolute precision less than 1 nano-Tesla at a sampling rate of 2.5 samples per second. The minute's mean value of $F$, its error mean value, the minute's standard deviation $SDF$, and its error were calculated, i.e., every 24 hours, 1440 quartets of data were recorded. Minute standard deviation of $F$ is defined as:

$$SDF_m = (1/N_m) \sum_{i=1}^{N_m} (F_i + F_{mean})^2 \qquad (1)$$

where $$F_{mean} = (1/N_m) \sum_{i=1}^{N_m} F_{mean} \qquad (2)$$

and $N_m$ is the number of samples per minute

### 3.3. The results: Geomagnetic quake as regional precursor for increasing regional seismicity

The connection between variations of local geomagnetic field and the earth currents was established in INRNE, BAS, Sofia, 2001 seminar [1]. The statistic of earthquakes that occurred in the region (1989- 2001), confirmed the Tamrazyan notes [2, 3], that the extreme tides are the earthquake's trigger. The Venedikov's code [4, 5] for calculating the regional tide force was used.

The signal for increasing regional seismic activity (incoming earthquake) is the "geomagnetic quake", which is defined as a jump (positive derivative) of daily averaged *SDF*. Such approach permits to compare by numbers the daily behaviour of the geomagnetic field with other days.

The predicted earthquakes can be identified as a maximum of function

$$S_{ChtM} = 10^{1.5M+4.8} / (D + Depth + Dist\,an\,ce)^2 \; [\text{energy/km}^2] \tag{3}$$

where D=40 km is fit parameter, calculated for every occurred in the region earthquake.

It is very important to note that the time scale of one minute, the correlation between the time period of increasing regional seismic activity (incoming earthquake), and tide extrema, recognized of predicted earthquake was established using the Alexandrov's code REGN for solving the over determined nonlinear systems [6, 10]. The very big worthiness of Alexandrov's theory and code is possibility to choose between two functions, which describe the experimental data with the same hi-squared, the better one.

The following Figures 1 and 2 are illustrations of the day without and with a signal for a "geomagnetic quake", calculated for one geomagnetic component, using *SDF* of Sofia one component geomagnetic monitoring: http://theo.inrne.bas.bg/~mavrodi/ [11].

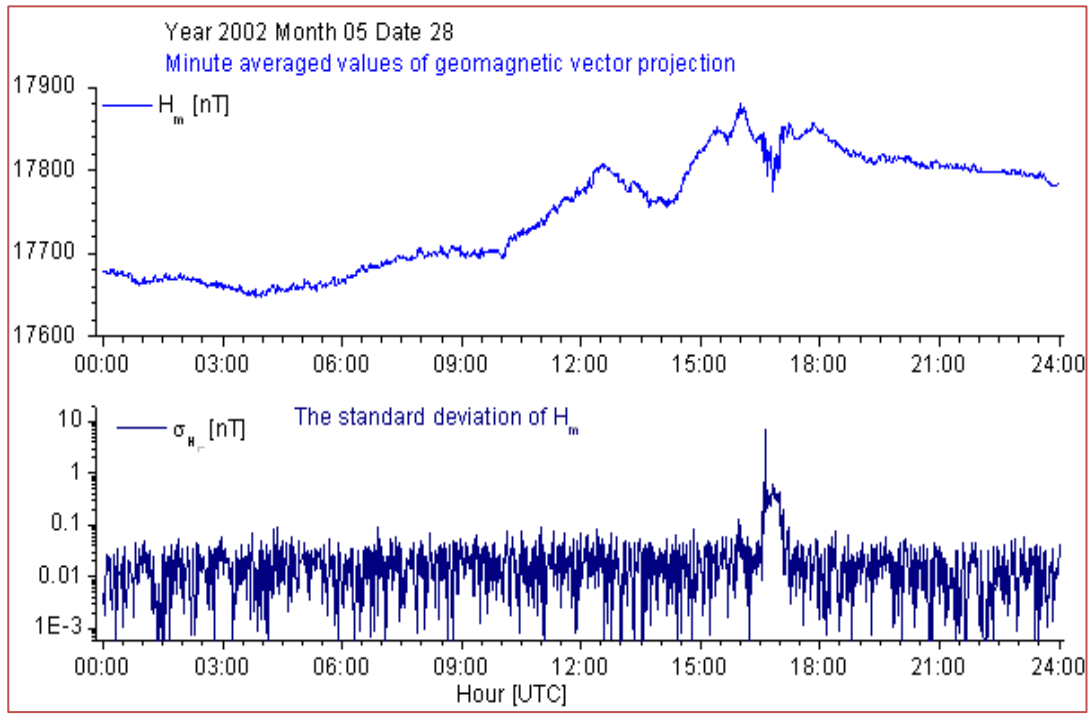

**Fig. 1.** Day with signal

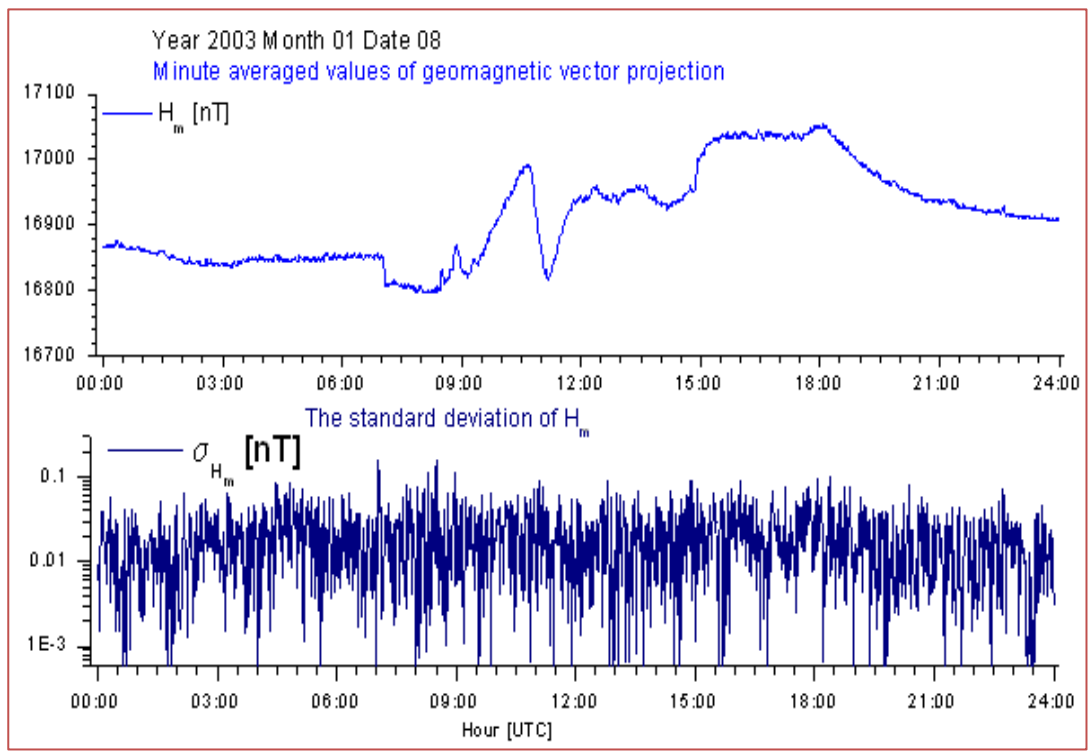

**Fig.2.** Day without signal

### 3.4. The results of geomagnetic vector monitoring [12-14]

In the following Figure 3 an example of the Skopje daily geomagnetic monitoring is presented:

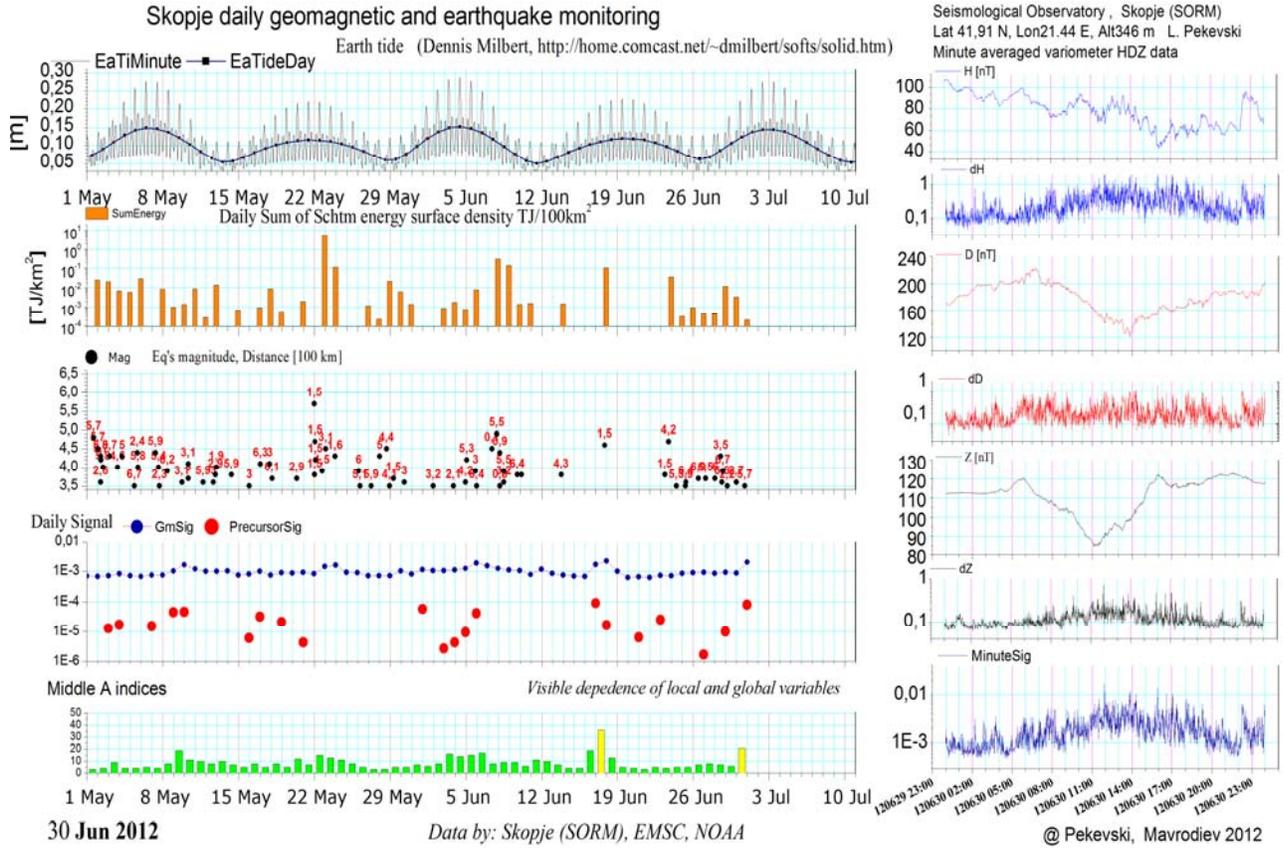

**Fig. 3**. Skopje daily geomagnetic and earthquake monitoring

In the right part of the figure are presented, up to down minute averaged values of *H, D, Z,* and its standard deviation of *SDH, SDD, SDZ* correspodingly, as well as the variable *MinuteSig,* measured with 10 samples per second.

$$MinuteSig = \left( SDH^2 + SDD^2 + SDZ^2 \right) / \left( H^2 + D^2 + Z^2 \right) \qquad (4)$$

In the left part of the figure, down to up are presented daily averaged values of Middle A indices (http://www.swpc.noaa.gov/alerts/k-index.html). Daily signal graph, where the *GmSig* is

$$GmSig = (1/1440)\sum_{1}^{1440} MinuteSig \qquad (5)$$

$$PrecursorSig = (1/2)(GmSig_{Today} - GmSig_{Yesterday}) / (A_{Today} - A_{Yesterday}) \quad (6)$$

In the next upper graph are presented the occurred earthquakes with index, which is the distance (<= 700 km) from the monitoring point. The next graph presents the daily sum of function *Schtm*, and the upper graphs presents the behaviour of minutes and daily averaged values of Earth's surface tidal movement, calculated with Dennis Milbert's code http://home.comcast.net/~dmilbert/softs/solid.htm.

In the next figure 4 is presented an example of the Panagjuriste INTERMAGNET daily geomagnetic monitoring with minute *XYZ* data hour calculations correspondingly.

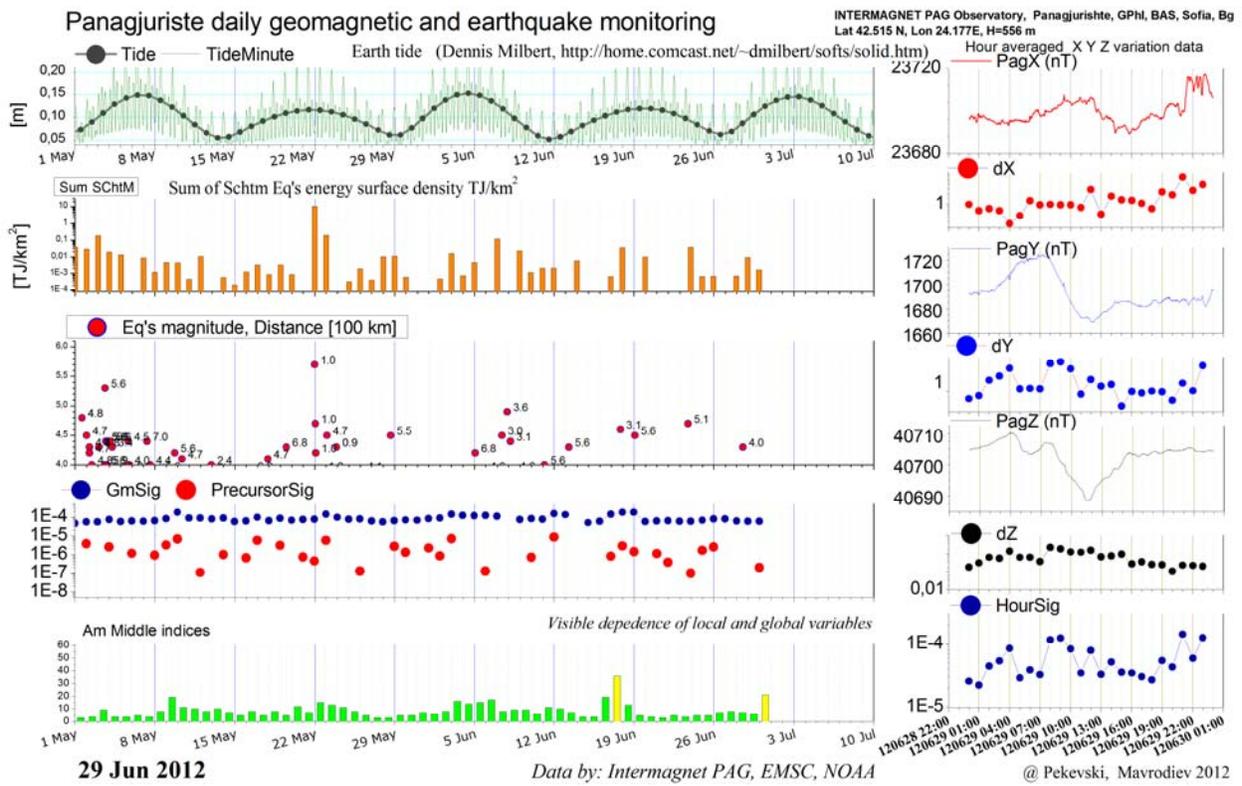

**Fig. 4**. Panagjuriste daily geomagnetic and earthquake monitoring

The above Skopje and Panagjuriste figures illustrate how our approach for Mw 5.6, Pernik earthquakes, May 22, 2012, 00:00 UTC time works, notwithstanding that Skopje data are with 10 samples per seconds and difference in distances.

Such posteriori analysis on the basis of INTERMAGNET for England, Alaska, India, Kamchatka, Hokkaido and other regions, where big earthquakes have occured also confirmed that the geomagnetic quake is a reliable regional precursor for imminent seismic increasing activity.

### 3.5 Some world earthquake statistics

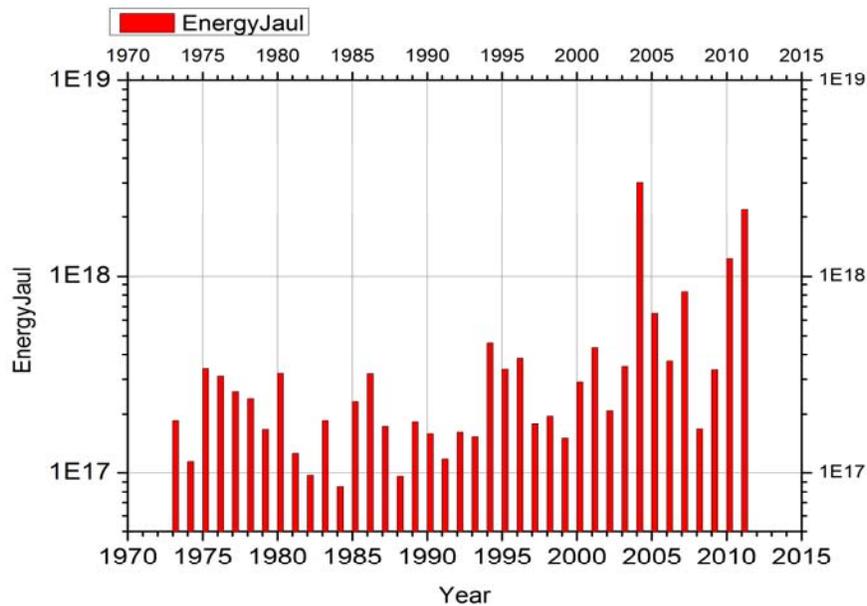

**Fig. 5**. The total World earthquakes energy with magnitude greater than 4 for the period of 1973-2011.

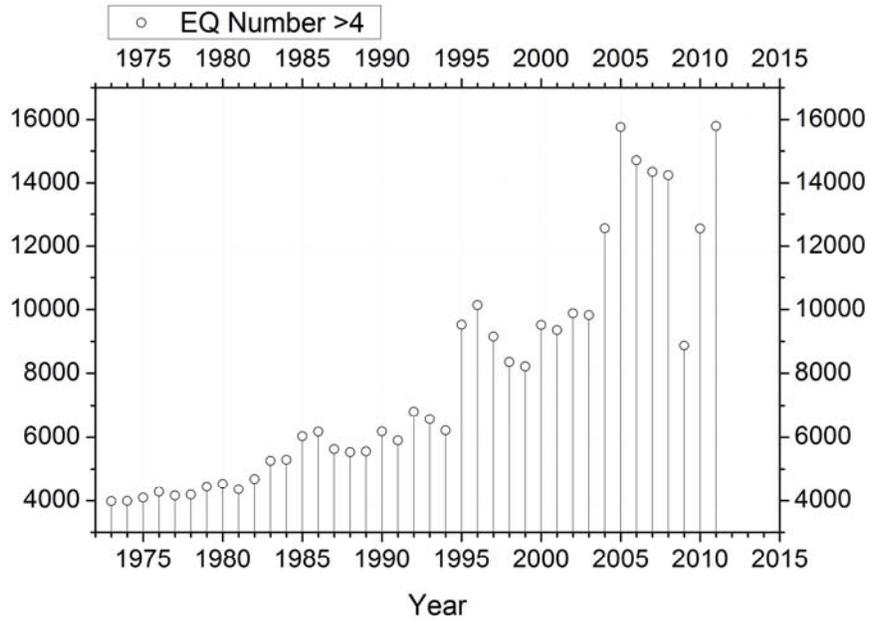

**Fig. 6.** The number of World earthquakes energy with magnitude greater than 4 for the period of 1973- 2011.

The earthquakes' statistic from 1982, presented in the above Figures 5, 6 and occured big Natural catastrophes illustrates our 2004 statement that our Civilization has some 5-7 years to solve "when, where and how" earthquakes' prediction problem http://theo.inrne.bas.bg/~mavrodi/.

The next  Figure 7 presents the statistical evidence that the dayli averaged Earth's local tides extremum are trigger of earthquakes (M>4).

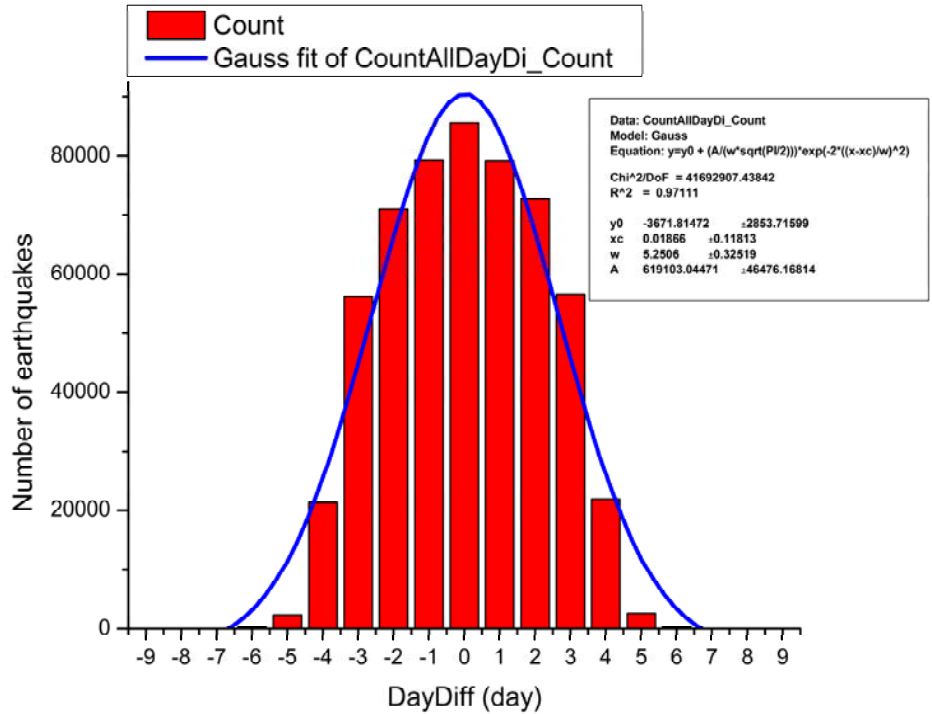

**Fig. 7**. The distribution difference between the time of local extreme tide and the time of occurred earthquake with magnitude greater than 4. The http://www.isc.ac.uk/ database was used.

### 3.6. Deep-focus earthquakes anomalous statistics, the axion-georeactor hypothesis for connection between seismicity and climate change [15]

The NEIC data analysis shows that the dynamics of spatial deep-focus earthquake distribution in the Earth's interior from 1973-2008 is characterized by the clearly defined periodical fine discrete structure with $L$=50 km, which is solely generated by earthquakes with magnitude M$\in$[3.9; 5.3] and only in the convergent boundary of plates. The explanation is based on an old (1774 year) paper of L. Euler et. al. Today authors give estimation of average values of stress in the upper mantle ($\sigma_0 \sim 1.8$ MPa), Young's modulus for the lithosphere ($E_{slab} \sim 55$ MPa) and upper mantle ($E_{mantle} \sim 1$ MPa). Figure 8 illustrates the dynamics of peak distribution for the period of 1972-2008. Figure 9 illustrates the greater Mag 5.3 earthquake distribution for all periods does not peak, which is in agreement with the understading of seismic proccesses concerning the possibility for big earthquake and Earth's core power.

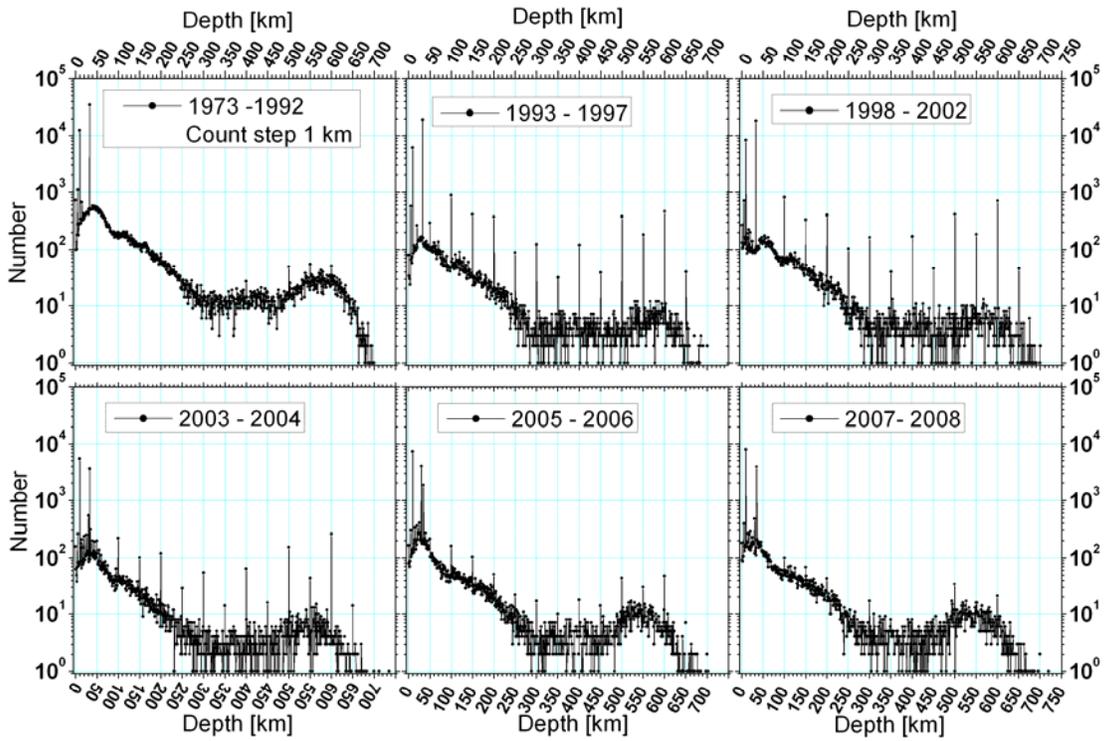

**Fig. 8** The years' dynamics of depth earthquake's number distributions

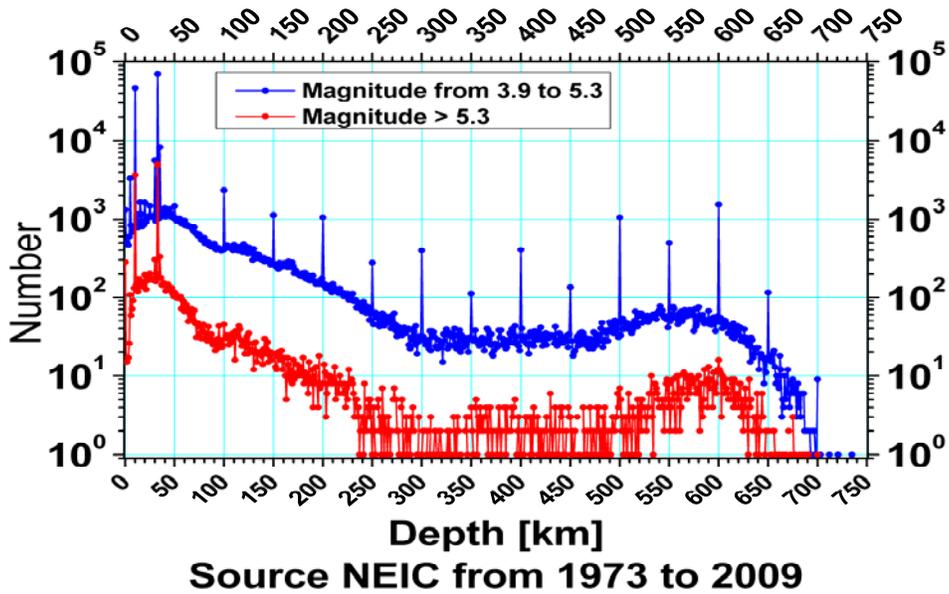

**Fig. 9.** The depth distribution of earthquake numbers for illustration of the fact that the magnitudes of earthquakes which occur in the sleb are limited from sleb's lift

In the next Figure 10, the spatial distributions of deep-focus earthquakes, with magnitude M∈[3.9, 5.3], shows the fine structure of depth distributions in deep-focus earthquakes over the years, 1973-2006. The map is created on the basis of data from http://pubs.usgs.gov/gip/dynamic/world_map.html.

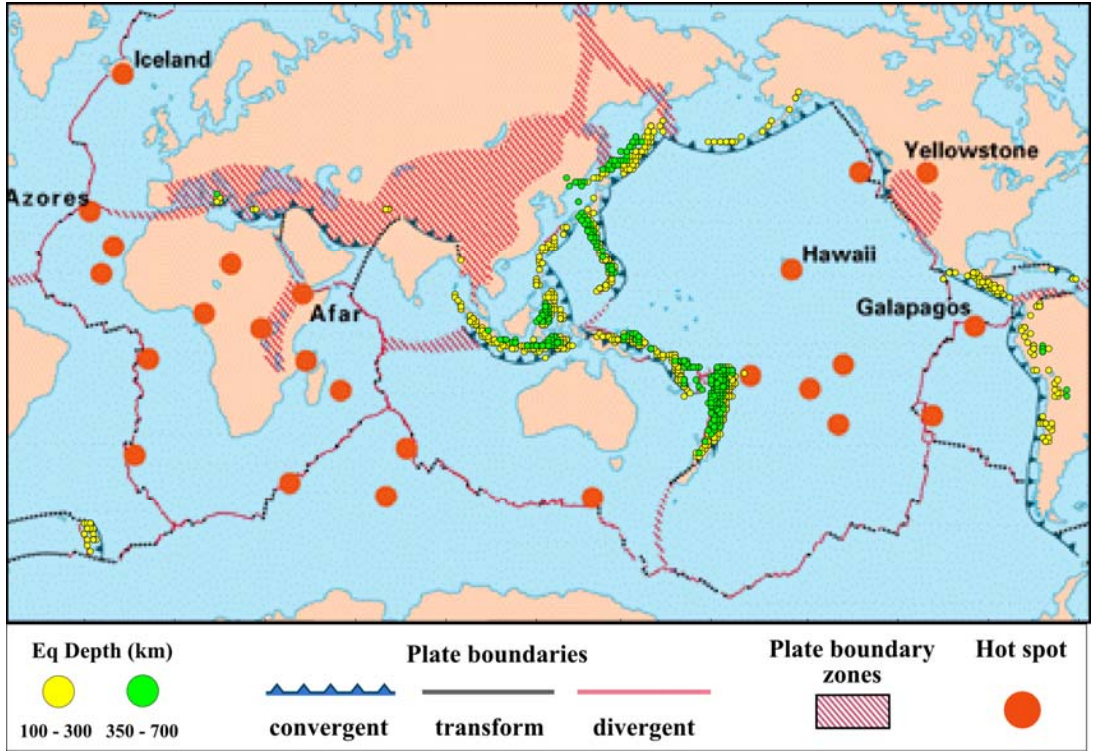

**Fig. 10**. The spatial distributions of deep-focus earthquakes with magnitude M∈[3.9, 5.3]

The existence of hot spots and ocean faults can be explained with the Sun-Earth interaction [15, 16] via axion and geo-reactors hypothesis.

One can estimate a strong negative correlation between the temporal variations of magnetic field steroidal component of the solar tachocline (the bottom of the convective zone) and the Earth's magnetic field (Y-component). Our first hypothesis is the possibility that hypothetical solar axions, which can transform into photons in external electric or magnetic fields (the inverse Primakoff effect), can be the instrument by which the magnetic field of convective zone of the Sun modulates the magnetic field of the Earth. The 57Fe solar axioms, as the main carriers of the solar-terrestrial connection, which by virtue of the inverse coherent Primakoff effect, can be transformed into photons in the iron Earth's core, increasing the temperature, as well as the effectiveness of the nuclear georeactors of Feoktistov type (second hypothesis) depending on the temperature. It is natural to propose that the hot spots and faults are image of Feoktistov's nuclear reactor distribution in the Earth's iron core canyons. In such ways can explain naturally the Vagener movements of continetal plates, the seismicity and climate behavior of the Earth.

Of course, creating the balance models, discovering of axions, understanding the proccesses in the Sun are the fundamental complex interdisciplinary scientific problems with greater significance. This can be compared with the process of Nuclear Disarments in 70-th, which was the consequence of the created Nuclear Winter Model.

## 4. Conclusions

The described above monitoring system now is developing for regional application with hope that in the region will be created a Center for data asquision system with the help of EU and NATO. It is important to stress that the solution of 'when, where, and how' earthquake's prediction problem is connected with meteorological hazards as well as with better step by step understanding the reasons for Earth seismicity and Climate change.

## Achnowledgements

I would like to note the enormous work of my Colleagues in the time of preparing the project. Of course, we are also very thankful to European Commission for financial support of our Project as well to NATO and organizers of NATO Advanced Research Workshop "The Black Sea: Strategy for Addressing its Energy Resource Development and Hydrogen Energy Problems", 7 – 10 October 2012, Batumi, Georgia for invitation. We hope that step by step such monitoring system of the Black Sea region will be created, which will help the central and local authorities to organize harmonic existence of the region.

**References**

[1]     Mavrodiev S.Cht., Thanassoulas C., "Possible correlation between electromagnetic earth fields and future earthquakes", ISBN 954-9820-05-X, Seminar proceedings, 23- 27 July, 2001, INRNE-BAS, Sofia, Bulgaria, 2001

[2]     Knopoff, L, Earth tides as a triggering mechanism for earthquakes, *Bull. Seism. Soc. Am.* 54**:**1865–1870, 1964

[3]     Tamrazyan G.P., Tide-forming forces and earthquakes, ICARUS, Vol.7, pp.59-65, 1967

[4]     [Tamrazyan G.P., Principal ragularities in the distribution of major earthquakes relative to Solar and Lubar tades and other Cosmic forces, ICARUS, Vol.9, pp.574-592, 1968

[5]     Venedikov, A. and Arnoso, R.: Program VAV/2000 for Tidal Analysis of Unevenly Spaced Data with Irregular Drift and Colored Noise, J. Geodetic Society of Japan, 47, 1, 281–286, 2001

[6]     Venedikov, A. P., Arnoso, R., and Vieira, R.: A program for tidal data processing, Computer & Geosciences, 29, 4, 487–502, 2003

[7]     Alexandrov L., Autoregularized Gauss-New-ton-Kantorovich iteration Process, Comm. JINR, P5-5515, Dubna, 1970

[8]     Alexandrov L., Regularized computational process of Newton-Kantorovich type, J. Comp. Math. And Math. Phys., 11, Vol. 1, 36-43, 1971

[9]     Alexandrov L., The program REGN (Regularized Gauss-Newton iteration method) for solving nonlinear systems of equations**,** Comm. JINR P5-7259, Dubna, 1973

[10]    Alexandrov L., Program code REGN, PSR 165 RSIK ORNL, Oak Ridge, Tennessee, USA,1983

[11]    Mavrodiev S. Cht., On the reliability of the geomagnetic quake as a short time earthquake's precursor for the Sofia region, Natural Hazards and Earth System Sciences (2004) 4: 433–447, SRef-ID: 1684-9981/nhess/2004-4-433

[12]    Mavrodiev, S. Cht., and Pekevski, L., Complex Regional Network for Earthquake Researching and Imminent Prediction, *Electromagnetic phenomena related to earthquakes and volcanoes,* Editor: Birbal Singh, Publ., Narosa Pub. House, New Delhi, pp. 135−146, 2008

[13]    Mavrodiev, S. Cht., Pekevski, L., and Jimseladze, T. Geomagnetic-Quake as Imminent Reliable Earthquake's Precursor: Starring Point for Future Complex Regional Network,

Electromagnetic phenomena related to earthquakes and volcanoes, Editor: Birbal Singh, Publ., Narosa Pub. House, New Delhi, pp. 116−134, 2008,

[14] Mavrodiev S. Cht., Pekevski L., On the Balkan- Black Sea- Caspian Complex NETWORK for Earthquake's Researching and Prediction, NATO Advanced research workshop management of urban earthquake risk in Central Asian ans Caucasus countries, Istanbul, 14-16 May, 2006

[15] Rusov, V. D.; Vaschenko, V. N.; Linnik, E. P.; Cht. Mavrodiev, S.; Zelentsova, T. N.; Pintelina, L.; Smolyar, V. P.; Pekevski, L., Mechanism of deep-focus earthquakes anomalous statistics, EPL (Europhysics Letters), Volume 91, Issue 2, pp. 29001 (2010)

[16] Rusov, V.D. et all, Solar Dynamo as host power pacemaker of Earth global climate, EU FP7 IRSES 2011 Project, "Complex research of Earthquake's Forecasting Possibilities, Seismicity and Climate Change Correlations", ISBN978-9989-631-04-7, Ohrid, Macedonia, BlackSeaHazNet Series, Volume 1, 2-5 May, 2011